\documentclass[reprint,amsmath,amssymb,aps]{revtex4-2}
\usepackage{graphicx}
\usepackage{dcolumn}
\usepackage{bm}
\usepackage{float}
\usepackage{amsmath}

\usepackage{xcolor}

\usepackage{comment}
\usepackage{mathrsfs}
\newcommand{\mycomment}[1]{}

\begin{document}

\title{Shear-Induced Collective Shape Oscillations in Dense Soft Suspensions}

\author{Ioannis Hadjifrangiskou}
\affiliation{Rudolf Peierls Centre for Theoretical Physics, University of Oxford,  Oxford OX1 3PU, United Kingdom}
\author{Rahil N. Valani}
\affiliation{Rudolf Peierls Centre for Theoretical Physics, University of Oxford,  Oxford OX1 3PU, United Kingdom}
\author{Diogo E. P. Pinto}\email{depinto@ciencias.ulisboa.pt}
\affiliation{Centro de Física Teórica e Computacional, Faculdade de Ciências, Universidade de Lisboa, P-1749-016 Lisboa, Portugal}
\affiliation{Departamento de Física, Faculdade de Ciências, Universidade de Lisboa, P-1749-016 Lisboa, Portugal}

\date{\today}

\begin{abstract}

Dense suspensions of deformable particles can exhibit rich nonequilibrium dynamics arising from complex flow–structure coupling. Using a multi–phase field model, we show that steady shear drives an initially disordered, dense, soft suspension into a positionally and orientationally ordered state, within which particles undergo robust self-sustained shape oscillations. These oscillations originate from repeated T1 neighbor exchanges that force the ordered particle lattice to cyclically traverse different ordered configurations, coupling particle deformation to evolving lattice topology. By identifying the lattice angle as a key variable, we construct a minimal one-degree-of-freedom model that quantitatively captures the limit cycle oscillation. Because these mechanisms rely only on deformability, packing, and shear, they provide a generic route to collective time-dependent behavior in dense soft suspensions.



\end{abstract}

\maketitle

\textit{Introduction} -- Dense suspensions of deformable particles, such as vesicles, capsules, droplets, and cells, exhibit rich nonequilibrium dynamics under applied external flows, that emerge from the coupling between deformability, particle-particle interactions, and confinement~\cite{Abkarian2002,Kantsler2006, Desmond2014, Brady1985, Mewis2012, Ness2022, Grassia2012, Larson1999, Princen2001}.
These soft-matter systems are ubiquitous in both nature and technology, from blood rheology and cellular transport to emulsions and microfluidic materials~\cite{Denkov2009, Cloitre2003, Hohler2005, Mason1995a, Becu2006, Mason1995b}. 
Such suspensions display nontrivial macroscopic responses including shear thinning \cite{Chen2010, Vermant2005} and thickening \cite{Sun2023, Wyart2014}, lane formation \cite{Sexton2011}, and flow-induced ordering that originate from microscopic shape fluctuations and rearrangements~\cite{Treado2021,Foglino2017}. 
Understanding such behavior is crucial for linking single-particle dynamics to collective behaviors in biological and synthetic suspensions.

At the single-particle level, the deformation and reorientation of soft objects in shear flow, exemplified by tank-treading, tumbling, and swinging motions, are now well characterized~\cite{Kantsler2006,Jeffery1922}. 
However, much less is known about how dense suspensions of deformable particles give rise to emergent time-dependent collective behavior when interparticle contacts and topological rearrangements dominate the dynamics.
The interplay between shape relaxation, crowding, and imposed shear can introduce a nonlinear feedback, potentially leading to novel dynamical regimes.

Here, we demonstrate that interactions among deformable particles in a simple shear flow give rise to a distinct collective state characterized by \textit{self-sustained shape oscillations}. Using a two-dimensional multi–phase field model~\cite{Cahn1958}, we show that these oscillations originate from contact-induced neighbor exchanges (T1 transitions) driven by differential 
flow between adjacent layers. We further develop a minimal dynamical system that captures the essential coupling between particle deformation and lattice geometry responsible for shape oscillations. Our results reveal a generic route to time-dependent collective behavior in soft suspensions, bridging microscopic rearrangements and macroscopic oscillatory responses.


\textit{Phase-field model} -  We use a 2D multi-phase field model where each particle is described by an independent scalar phase field, $\phi_i(\textbf{x})$, that continuously varies from $0$ to $1$ \cite{Mueller2019, Basan2013}. The motion of each phase field is governed by a local velocity field $\textbf{v}_i(\textbf{x})$ according to the equation of motion~\citep{supplementary_m},

\begin{equation}
\label{eqEOM}
    \partial_{t} \phi_i(\textbf{x}) + \textbf{v}_i(\textbf{x}) \cdot \nabla \phi_i(\textbf{x}) = -J_0 \frac{\delta \mathscr{F}}{\delta \phi_i(\textbf{x})} \\ ,
\end{equation}
 
\noindent where $\mathscr{F}$ is a free energy. The right-hand side of Eq.~\eqref{eqEOM} corresponds to the relaxation dynamics of the phase-fields to a free-energy minimum at a rate $J_0$. We impose the shear flow field by setting $\textbf{v}_i = \dot{\gamma}y\hat{\textbf{x}}$, where $\dot{\gamma}$ corresponds to the shear rate.

The shape of the particles is characterized by the rank $2$ tensor:

\begin{equation}
\label{Rtensor}
    \textbf{R}_i = - \int d\mathbf{x} \bigg[ \nabla \phi_i \nabla \phi_i^T \bigg] \\,
\end{equation}

\noindent so that the eigenvectors of this tensor, $\textbf{d}^{\parallel}$ and $\textbf{d}^{\bot}$, lie along and perpendicular to the elongation of the phase field, respectively. We also define the elongation of each phase field as $r= (\lambda_1/\lambda_2) - 1$, where $\lambda_i$ are the corresponding eigenvalues and their ratio gives the shape aspect ratio.

In the following, we consider a system of $N=100$ deformable particles, with a radius of $R=8$ in lattice units, which move in a thin channel of length $L_x = 70$ and width $L_y=240$, leading to a dense packing. The channel is periodic along the $x$-direction and the walls are enforced through a Neumann boundary condition ($\nabla\phi\cdot \hat{\textbf{y}} = 0$) at the top and bottom. We initialize the system by positioning the centroid of the phase fields randomly in the simulation box with a starting radius of $R/2$. We then let the system relax without imposing the shear flow for $20000$ time steps, after which we simulate with the imposed shear flow for another $10^{6}$ time-steps (see Supplementary Material~\citep{supplementary_m} for more details).


\textit{Shear induces bond and nematic order} - We start the simulation with random initial conditions, such that we are in a disordered state. The imposed shear drives particle motion, while free energy relaxation governs how the particles respond to this forcing. We observe that as $\dot{\gamma}$ increases, order emerges in the system. In particular, positional order emerges in the form of layers of particles moving coherently along the same direction (similar to shear thinning \cite{Chen2010, Vermant2005}), and nematic order emerges due to similar shape deformations of neighboring particles. We quantify the positional order through the global bond order parameter \cite{Steinhardt1983}:

\begin{equation}
    \Psi_6 = \bigg\langle \bigg\lvert \  \frac{1}{N} \sum_{j=1}^N \psi_{6,j} \bigg\rvert \bigg\rangle \,
\end{equation}

\noindent where the bond-orientational order $\psi_{6,j}=\frac{1}{N_{j,nn}}\sum_{k\in nn}e^{i6\theta_{jk}}$, the sum is over the nearest neighbors ($nn$) and $\theta_{jk}$ is the angle between the $x$-axis and the bond vector linking particles $j$ and $k$. We define nearest neighbors through the Delaunay triangulation of the phase-field centroids. For the nematic order, we use the order parameter \cite{Stephen1974}:

\begin{equation}
    \Psi_2^L = \bigg\langle \bigg\lvert \  \frac{1}{N} \sum_{j=1}^N \psi_{2,j} \bigg\rvert \bigg\rangle \,
\end{equation}

\noindent where $\psi_{2,j}=\frac{1}{N_{j,nn}}\sum_{k\in nn}e^{i6\theta_{jk}^d}$, the sum is over the nearest neighbors and $\theta_{jk}^d$ is the angle between the deformation axes of particles $j$ and $k$.

\begin{figure}[t]
\includegraphics[width=1\columnwidth]{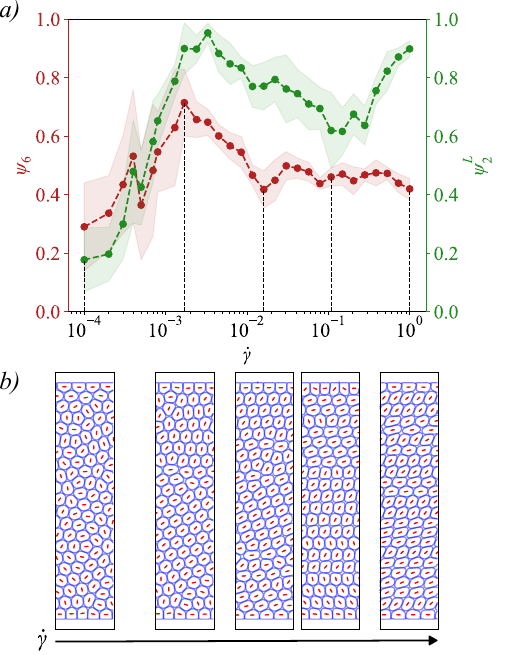}
\caption{\textbf{Order emerges from induced shear flow}. a) Bond and nematic order as a function of shear rate. The order parameters were calculated using the last $50$ time frames of the simulations, each with a total of $250$ time frames. These results were averaged over $10$ different samples. The shaded region represents the standard deviation of this average. b) Schematic representations of the phase field simulations, where the axis of elongation is placed at the center of each phase field in red. Each schematic corresponds to one of the vertical dashed lines in panel a).
}
\label{Fig: Order}
\end{figure}

Figure~\ref{Fig: Order}a) shows a non-monotonic dependence of the bond and nematic order on the applied shear rate, $\dot{\gamma}$. At low shear rates, the imposed flow is too weak to induce particle motion, and the system remains disordered (first schematic in Fig.~\ref{Fig: Order}b and Supplemental Video S1~\citep{supplementary_m}). 

Increasing $\dot{\gamma}$ leads to relative particle motion and neighbor exchanges (T1 topological transitions), marking a transition to a flowing state with similarities to a yield-stress transition observed in forced phase-field systems and other yield-stress materials~\citep{doi:10.1126/sciadv.adf8106,Bonn2017}. An initial peak in both order parameters (second schematic in Fig.~\ref{Fig: Order}b and Supplemental Video S2~\citep{supplementary_m}) appears near the onset of T1 topological transitions~\cite{Princen1983}, accompanied by large fluctuations around the mean value. As T1 events become more frequent, particles align and form correlated blocks that deform collectively along the principal strain direction, increasing both bond and orientational order.

At higher shear rates, frequent T1 transitions allow particles to slip past one another, leading to the formation of single-particle-wide lanes along the flow direction (third schematic in Fig.~\ref{Fig: Order}b and Supplemental Video S3~\citep{supplementary_m}). Given this coherent motion, the T1 transitions happen predominantly along the horizontal direction as lanes move relative to each other and are accompanied by shape oscillations as phase fields are deformed due to shear flow and rotated due to steric interactions. The reduction of correlated translational motion causes the bond order to decrease, while emergent out-of-phase shape oscillations between particles in neighboring lanes reduce the nematic order (fourth schematic in Fig.~\ref{Fig: Order}b and Supplemental Video S4~\citep{supplementary_m}).

At very high shear rates, the lane structure stabilizes and bond order saturates, while the amplitude of shape oscillations diminishes. Particles increasingly align with the flow strain direction ($\theta=\pi/4$), leading to a renewed increase in nematic order (fifth schematic in Fig.~\ref{Fig: Order}b and Supplemental Video S5~\citep{supplementary_m}).

\textit{Sheared configurations induce shape oscillations} - Shape oscillations in our system emerge from collective dynamics under external shear. An isolated particle subjected to shear deforms along the flow strain axis but does not oscillate (see Supplementary Material~\citep{supplementary_m}), with its elongation set by the imposed shear rate and material parameters. To isolate the mechanism underlying the oscillations, we therefore consider an ordered regime, initializing particles on an ideal hexagonal lattice and applying shear until collective shape oscillations develop.

\begin{figure}
\centering
\includegraphics[width=1\columnwidth]{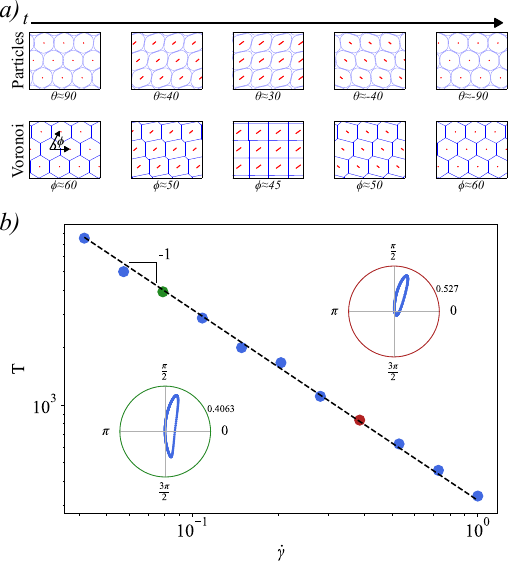}
\caption{\textbf{Shape oscillations in the phase field model}. a) Snapshots of phase field configurations (top) and corresponding Voronoi constructions (bottom) at different times during one period of shape oscillation. For each, the angle of the orientation ($\theta$) and the lattice angle ($\phi$), taken from the Delaunay neighbors of each phase field, are shown, for different points in time. b) Period of shape oscillations as a function of the shear rate for the phase field model with regular initial conditions. The axis are in logscale and a line with slope one is drawn, highlighting the linear dependence of the period with shear rate. Two limit cycles in polar coordinates are shown which correspond to $\dot{\gamma}=0.0778$ and $\dot{\gamma}=0.3857$ (green and red respectively). The radial coordinate corresponds to the elongation of particles $r$, and the polar angle to their nematic director from the $x$-axis ($2\theta$).}
\label{Fig: T1}
\end{figure}

Figure~\ref{Fig: T1}a) illustrates representative phase-field configurations with lattice angle $\phi$ over one full shape-oscillation cycle just after the onset of oscillations, together with the corresponding particle orientation angles $\theta$. Starting from the initial $\theta=\pi/2$ elongation, particles elongate and rotate toward $\pi/4$, the direction of the strain rate imposed by simple shear. Because different layers translate at different velocities, layers eventually overtake one another, leading to neighbor exchanges (T1 transitions). Immediately after a T1 event, the particle above blocks further deformation along $\pi/4$, forcing particle $i$ to reorient in order to relax the rearrangement and fill the newly available space. Consequently, the deformation axis rapidly switches from $\pi/4$ to $-\pi/4$. The imposed flow then restores the initial configuration, and the oscillatory cycle repeats (see Supplemental Video S6 and S7~\citep{supplementary_m}). A limit-cycle representation of this motion (green) in the polar phase space formed by particle elongation $r$ and its nematic director angle $2\theta$ is shown in Fig.~\ref{Fig: T1}b).


If the shear rate $\dot{\gamma}$ is sufficiently large that particles cannot fully relax between successive T1 events, full oscillations no longer occur (see Supplemental Video S8~\citep{supplementary_m}). This is evident in the red $r$–$2\theta$ limit cycle in Fig.~\ref{Fig: T1}b), where the phase-space trajectory rapidly traverses negative orientation angles. At even higher shear rates, the limit cycle becomes confined to the first quadrant of the polar phase space. In this partial-oscillation regime, particle orientations approach ever more closely the strain direction of the imposed shear flow ($\theta = \pi/4$).

In Fig.~\ref{Fig: T1}b), we show that the period of shape oscillations decreases linearly with increasing shear rate, confirming that the oscillations are mediated by the applied shear. We note that although the amplitude of the shape oscillations decreases with shear rate, the sequence of configurations traversed by the particles remains unchanged. As shown in Fig.~\ref{Fig: T1}a), when the initial condition corresponds to a hexagonal lattice and the T1 events occur at a square lattice configuration (where each particle has four neighbors), the intermediate states consist of oblique lattices with varying orientations that depend on the degree of particle shape elongation. The rate at which these configurations are traversed is therefore set by the shear rate. In contrast, the amplitude of the shape oscillations is governed by the ratio of the shear rate to the free-energy relaxation rate ($J_0$). Consequently, at high shear rates, although the system continues to explore the full range of configurations, albeit more rapidly, the shape oscillations become only partial, as the particles do not have sufficient time to fully relax between successive configurational transitions.

\begin{figure*}
    \centering
    \includegraphics[width=0.9\linewidth]{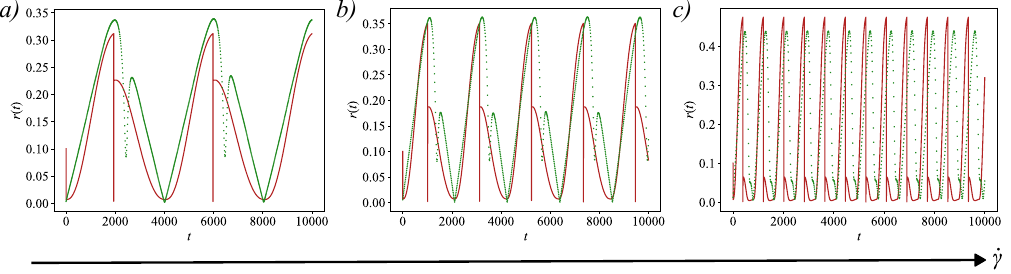}
    \caption{\textbf{Comparison of shape oscillations between models with increasing shear rate}.
    Plots of the elongation, $r(t)$ for different values of the shear rate. $\dot{\gamma} = (0.0778, 0.1487, 0.3857)$ for $a), b)$ and $c)$ respectively. Red lines represent the solutions calculated numerically from the minimal model in Eq.~\eqref{r:eqn}, while the green dots correspond to the results from phase-field simulations.} 
    \label{fig:ODE_plots}
\end{figure*}

\textit{Minimal dynamical model} - For the system dynamics with regular initial conditions described above, we find that the shape oscillations are driven by the lattice configuration angle $\phi \in [\pi/4, \pi/3]$ (see Fig.~\ref{Fig: T1} a))~\footnote{This angle is calculated by doing a bond-orientational order analysis between the phase-fields' center of mass and extracting the dominant bond angles. We use the periodic Delaunay triangulation to define the neighbors of a given phase field.}. 
To approximate the periodic nature of the series of T1 transitions \cite{Princen1983}, we fit a simple analytical form for $\phi(t)$ to the simulations giving
\begin{align}
    \phi(t) = \dfrac{\pi}{12}\cos^{2}\left({\omega t}\right) + \dfrac{\pi}{4},
\end{align}
where $\omega = k \dot{\gamma}$, $\dot{\gamma}$ is the imposed shear rate and $k$ is a fitting parameter that is determined by system specific parameters, such as the radius of particles. We found that $k = 10^{-2}$ is a good fit for all values of $\dot{\gamma}$ of interest (see Supplementary Material \citep{supplementary_m}).

We can now formulate a simple evolution equation for the particle orientation angle $\theta$ that preserves the oscillation period and captures the key qualitative features of the shape dynamics:
\begin{align} \label{theta_eqn}
\dot{\theta}(t) = -3\,|\dot{\phi}| = -\dfrac{\pi \omega}{4}\,|\sin(2\omega t)|.
\end{align}
In Eq.~\eqref{theta_eqn}, the negative sign and absolute value reflect that phase field rotation is always clockwise. The prefactor is chosen so that $\theta(t)$ and $\phi(t)$ share the same period, while accounting for the nematic symmetry of $\theta$. We further impose that whenever $\phi(t)=\pi/4$, corresponding to a square lattice and a T1 transition, the orientation angle $\theta$ is reset to $-\pi/4$. Together, this lead to a piecewise solution for $\theta(t)$ (see Supplementary Material~\citep{supplementary_m}).

We now formulate a dynamical equation governing the evolution of the particle shape, $r(t)$, defined as $r = AR - 1$, where $AR$ denotes the aspect ratio of each particle:
\begin{align} \label{r:eqn}
\dot{r}(t) = \dot{\gamma}(r + 1)\sin(2\theta) + A (\pi/3 - \phi) (r + 1) - \epsilon r^{3} + \frac{K}{r},
\end{align}
where $\dot{\gamma}$ is the imposed shear rate and $A$, $K$, and $\epsilon$ are positive constants.


The first term captures shear-induced elongation or contraction for positive or negative $\theta$, respectively. The second term encodes the effect of lattice geometry, with $\phi=\pi/4$ (square lattice) maximizing free space and promoting elongation, relative to $\phi=\pi/3$ (hexagonal lattice). The third term represents the free energy cost of deforming particles away from their isotropic equilibrium shape \footnote{This choice follows the phase-field free-energy formulation, in which gradients in the phase field are penalized—more elongated particles exhibit sharper gradients at the apoapsis.}. The final term ensures numerical stability by preventing strictly isotropic shapes; $K$ is chosen sufficiently small so as not to affect the results. \\
Fig.~\ref{fig:ODE_plots} compares the time evolution of the elongation $r(t)$ obtained from the minimal model and the full phase-field simulations for increasing shear rates $\dot{\gamma}$ (with parameters $A=2$, $\epsilon=2.3$, $K=0.01$). The shape oscillations display a pronounced asymmetry, characterized by a large primary peak which is immediately followed by a smaller secondary peak within each cycle. In the model, this asymmetry arises from the orientation-dependent shear term, which enhances elongation when particles are favorably aligned with the flow and promotes contraction when the orientation is reversed. As the shear rate increases, the amplitude of the secondary peak decreases due to the reduced time available for free-energy relaxation between successive configurational transitions, while the primary peak increases slightly as a result of enhanced shear-induced stretching. Despite its drastic reduction in complexity, the minimal model quantitatively reproduces these features, indicating that the observed periodic shape oscillations originate from the competition between shear-induced deformation, thermodynamic relaxation, and inter-particle interactions manifested through topological (T1) transitions.

\textit{Conclusions} -- We showed that steady shear organizes dense suspensions of deformable particles into translating lanes with bond and orientational order, within which collective, self-sustained shape oscillations emerge. These oscillations, absent for isolated particles, result from repeated T1 neighbor exchanges that couple particle deformation to lattice geometry. Identifying the lattice angle as the key geometric variable enabled a minimal one-degree-of-freedom model of the aspect ratio that quantitatively captures the oscillatory regimes and reveals the underlying mechanism as a competition between flow-induced stretching, free-energy relaxation, and topological rearrangements mediated by T1 transitions.

Although we focused on shear flow, the mechanism should apply to any flow with velocity gradients, as illustrated for Poiseuille flow in the Supplementary Material~\citep{supplementary_m}. Because these ingredients are common to dense soft suspensions such as emulsions, vesicles, microcapsules, and confluent cells, we expect similar oscillatory dynamics to be experimentally accessible and to have measurable rheological consequences.

\textit{Acknowledgments} -- I.H. acknowledges the support of the Gould \& Watson Scholarship. R.V. acknowledges the support of the Leverhulme Trust [Grant No. LIP-2020-014] and the ERC Advanced Grant ActBio (funded as UKRI Frontier Research Grant EP/Y033981/1). DEPP acknowledges support from the UKRI Horizon Europe Guarantee MSCA Postdoctoral Fellowship No. EP/Z002761/1 and the financial support from the Portuguese Foundation for Science and Technology under contracts: UID/00618/2023 and 2024.15260.PEX.

\section*{Supplemental Material}

\subsection{Model details}

We simulate Eq.~(1) in the main text using a finite-difference scheme on a square grid with a predictor-corrector step. We also use a space decomposition method, as in \cite{Mueller2019}, where each phase field is only solved inside a subdomain that covers an area of $30\times 30$ grid sites, centred on the centre of mass of the phase field. This subdomain is variable after the equilibration protocol. We set a threshold value of $\phi_i(\textbf{x})=0.01$, and if the distance from \textbf{x} to the sides of the subdomain is smaller than $4$ grid sites, then we increase the subdomain size. On the other hand, if it is larger, we decrease it. We also do not enforce a square subdomain. This optimization helps deal with highly elongated cells without fixing a large subdomain from the start.

The free energy term in Eq.~(1) of the main text defines the dynamics of the individual interfaces and is written as $\mathscr{F} = \mathscr{F}_{CH} + \mathscr{F}_{area} + \mathscr{F}_{rep}$, where

\begin{equation}
\label{CH}
    \mathscr{F}_{CH} = \sum_i \frac{\gamma}{\lambda} \int d\textbf{x} \big[ 4\phi_i^2(\textbf{x}) (1-\phi_i(\textbf{x}))^2 + \lambda^2 (\nabla \phi_i(\textbf{x}))^2  \big] \\,
 \end{equation}

\begin{equation}
\label{area}
    \mathscr{F}_{area} = \sum_i \mu \bigg[ 1 - \frac{1}{\pi R^2} \int d\textbf{x} \phi_i^2(\textbf{x})   \bigg] \\,
 \end{equation} 

\begin{equation}
\label{rep}
    \mathscr{F}_{rep} = \sum_i \sum_{j\neq i} \frac{\kappa}{\lambda} \int d\textbf{x} \phi_i^2(\textbf{x}) \phi_j^2(\textbf{x}) \\.
\end{equation}

\noindent Equation \eqref{CH} is a Cahn-Hilliard free energy that encourages $\phi_i(\textbf{x})$ to be equal to $1$ or $0$ inside or outside phase field $i$, respectively. The phase field boundary is located at $\phi_i(\textbf{x}) = 1/2$ and has width $\sim 2\lambda$, set by the gradient term. $\gamma / \lambda$ sets an energy scale. This term controls the deformability of the individual phase fields. Compressibility is described by Eq.~(\ref{area}), which imposes a soft constraint, of strength $\mu$, restricting the area of each phase field to $\pi R^2$. Equation~(\ref{rep}) penalizes overlap between phase fields with an energy scale $\kappa / \lambda$.

We simulate the dynamics with particles of radius $R = 8$ in a wide channel. Initially, cells with radius $R/2$ are placed randomly. They are then relaxed for $20000$ time-steps without external flow. The simulations are then run for $1000000$ time-steps. The grid spacing is set to $\Delta x = 1$ and the interval between time-steps is set to $\Delta t = 0.1$. Parameter values are $\lambda = 2.0$, $\gamma = 0.06$, $\kappa = 0.5$, $\mu = 20$, and $J_0 = 0.5$.

\subsection{Single phase field under shear}

\begin{figure}
\centering
\includegraphics[width=1\columnwidth]{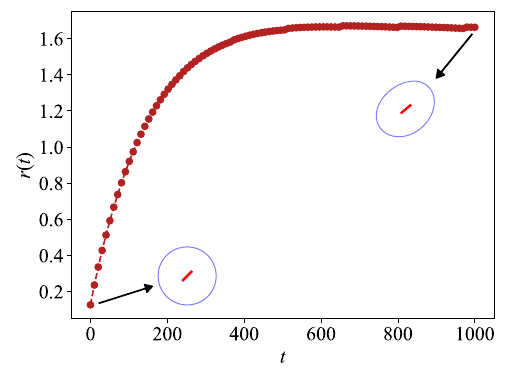}
\caption{\textbf{Single phase field under shear flow.} Elongation of the phase field as a function of time for shear rate, $\dot{\gamma}=1$.}
\label{Fig: SM1}
\end{figure}

In the case of a single phase field under the linear shear profile, the particle is deformed due to the gradients in velocity but it does not oscillates. Fig.~\ref{Fig: SM1} shows a time series of the elongation of a single phase field. This shows that the oscillatory regime is a collective effect due to multiple interacting deformable phase fields.

\subsection{Fitting lattice configurations ($\phi$)}
\begin{figure}
\centering
\includegraphics[width=1\columnwidth]{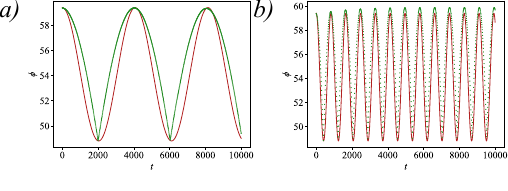}
\caption{\textbf{Lattice configuration fits to simulation data.} Fitting curves (red) of the function $\phi=A\cos^2(kt)$ to simulations data (green), for $A=10.6$. In a) fitting is shown for $\dot{\gamma}=0.0778$ and in b) for $\dot{\gamma}=0.3857$. The fitting parameter $k=0.01$ is independent of shear rate.}
\label{Fig: SM2}
\end{figure}

In Fig.~\ref{Fig: SM2} we show the empirical fit done to the lattice angle $\phi$. We fit a function of the form, $\phi=A\cos^2(kt)$, where $A$ represents the amplitude of the oscillation and $k$ is a fitting parameter. We find $k=0.01$ to be a good fit for all shear rates explored.

\subsection{Piecewise solution for $\theta(t)$}
Recall the time evolution of the particle orientation angle, $\theta$, given by Eq.~\eqref{theta_eqn}
\begin{align}
\dot{\theta}(t) = -3\,|\dot{\phi}| = -\dfrac{\pi \omega}{4}\,|\sin(2\omega t)|.
\end{align}

Defining $\tau_{n} = \dfrac{(2n+1)\pi}{2\omega}$ as the times where a T1 happens, and $\theta(\tau_{n}^{+}) = -\pi/4$, for an initial condition $\theta(0) = \pi/2$ which corresponds to $\phi(0) = \pi/3$, we may define $\theta(t)$ in a piecewise fashion:

\begin{equation}
\theta(t) =
\begin{cases}
\displaystyle
\frac{\pi}{2}
- \frac{\pi\omega}{4}
  \int_{0}^{t}\! \big|\sin(2\omega s)\big|\,\mathrm{d}s,
& t \in [0, \tau_0), \\[12pt]
\displaystyle
-\frac{\pi}{4}
- \frac{\pi\omega}{4}
  \int_{\tau_n}^{t}\! \big|\sin(2\omega s)\big|\,\mathrm{d}s,
& t \in [\tau_n, \tau_{n+1}),\quad 
\end{cases}
\qquad
\label{eq:theta_piecewise}
\end{equation}
where it is understood that $\theta + \pi \equiv \theta$ due to its nematic nature. In writing this down, we are approximating the relatively fast timescale of the T1 as instantaneous. \\ 

\subsection{Poiseuille flow}

\begin{figure}
\centering
\includegraphics[width=1\columnwidth]{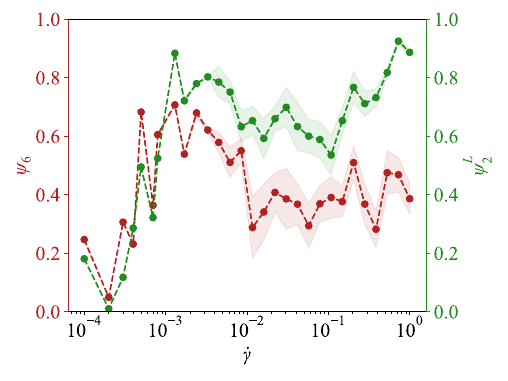}
\caption{\textbf{Order parameters under Poiseuille flow.} Bond and nematic order parameters as a function of the maximum velocity of a quadratic Poiseuille flow, $\dot{\gamma}$. Similar to Fig.~1 in the main text, translational and orientational order emerges due to the collective interactions between particles and the flow. The order parameters were calculated using the last $50$ time frames of the simulations, each with a total of $250$ time frames. These results were averaged over $3$ different samples. The shaded region represents the standard deviation of this average.}
\label{Fig: SM3}
\end{figure}

In the main text we explored the case where the dense suspension is under linear shear flow, but the results shown are general, as long as, there are gradients in the velocity field. To corroborate this, we simulated the same system but under a parabolic Poiseuille flow. In Fig.~\ref{Fig: SM3}, we show the bond-order parameter and the nematic order parameter, as in Eqs.~(3) and (4), for different maximum velocities $\dot{\gamma}$ of the flow profile. We recover a similar result to the one shown in Fig.~1, highlighting the generality of the physical behavior explored in this work.

\subsection{Captions for Supplemental Videos}

\textbf{Supplemental Video S1}: Video showing the time evolution of the system for $\dot{\gamma}=0.0001$. This video accompanies the first schematic of Fig.~1b.

\textbf{Supplemental Video S2}: Video showing the time evolution of the system for $\dot{\gamma}=0.0017$. This video accompanies the second schematic of Fig.~1b.

\textbf{Supplemental Video S3}: Video showing the time evolution of the system for $\dot{\gamma}=0.0161$. This video accompanies the third schematic of Fig.~1b. 

\textbf{Supplemental Video S4}: Video showing the time evolution of the system for $\dot{\gamma}=0.1083$. This video accompanies the fourth schematic of Fig.~1b. The shear profile is $\textbf{v}_i = \dot{\gamma}y\hat{\textbf{x}}$, the lanes moving backwards in these videos are an artifact due to the chosen frame rate.

\textbf{Supplemental Video S5}: Video showing the time evolution of the system for $\dot{\gamma}=1$. This video accompanies the fifth schematic of Fig.~1b. The shear profile is $\textbf{v}_i = \dot{\gamma}y\hat{\textbf{x}}$, the lanes moving backwards in these videos are an artifact due to the chosen frame rate.

\textbf{Supplemental Video S6}: Video showing the time evolution of a lane of particles for $\dot{\gamma}=0.1487$ when it is initialized from a hexagonal lattice configuration. This video accompanies the schematics in Fig.~2a.

\textbf{Supplemental Video S7}: Video showing the time evolution of the Voronoi tesselation of the centroids of the phase fields for $\dot{\gamma}=0.1487$ when it is initialized from a hexagonal lattice configuration. This video accompanies the schematics in Fig.~2a.

\textbf{Supplemental Video S8}: Video showing the time evolution of a lane of particles for $\dot{\gamma}=0.3857$ when it is initialized from a hexagonal lattice configuration. This video accompanies the high shear polar plot in Fig.~2b.


%

-----------------------------------------

\end{document}